\documentclass[aps,superscriptaddress,preprint]{revtex4}%
\usepackage{mathrsfs}
\usepackage{amsfonts}
\usepackage{amsmath}
\usepackage{amssymb}
\usepackage{graphicx}
\usepackage{diagbox}
\usepackage{booktabs}
\usepackage{epstopdf}
\usepackage{graphicx}
\usepackage{mathrsfs}
\usepackage{bm}
\usepackage{dcolumn}
\usepackage{epstopdf}
\usepackage{dsfont}
\usepackage{tabularx}
\usepackage{float}
\usepackage{color}
\usepackage{pifont}
\usepackage{siunitx}
\usepackage{multirow}%
\providecommand{\U}[1]{\protect\rule{.1in}{.1in}}

\begin{document}
\title{Three-gap superconductivity with $T_{c}$ above 80 K in hydrogenated 2D monolayer LiBC}

\author{Hao-Dong Liu}
\affiliation{Institute of Applied Physics and Computational Mathematics and National Key Laboratory of Computational Physics, Beijing 100088, China}
\affiliation{Graduate School, China Academy of Engineering Physics, Beijing 100088, China}

\author{Bao-Tian Wang}
\affiliation{Institute of High Energy Physics, Chinese Academy of Sciences, Beijing 100049, China}
\affiliation{Spallation Neutron Source Science Center, Dongguan 523803, China}

\author{Zhen-Guo Fu}
\thanks{E-mail: fu\_zhenguo@iapcm.ac.cn}
\affiliation{Institute of Applied Physics and Computational Mathematics and National Key Laboratory of Computational Physics, Beijing 100088, China}

\author{Hong-Yan Lu}
\thanks{E-mail: hylu@qfnu.edu.cn}
\affiliation{School of Physics and Physical Engineering, Qufu Normal University, Qufu 273165, China}

\author{Ping Zhang}
\thanks{E-mail: zhang\_ping@iapcm.ac.cn}
\affiliation{Institute of Applied Physics and Computational Mathematics and National Key Laboratory of Computational Physics, Beijing 100088, China}
\affiliation{School of Physics and Physical Engineering, Qufu Normal University, Qufu 273165, China}

\date{\today}

\begin{abstract}
Although the metalization of semiconductor bulk LiBC has been experimentally achieved, various flaws, including the strong lattice distortion, the uncontrollability of phase transition under pressure, usually appear. In this work, based on the first-principles calculations, we propose a new way of hydrogenation to realize metalization. Using the fully anisotropic Migdal-Eliashberg theory, we investigate the superconducting behaviors in the stable monolayers LiBCH and LiCBH, in which C and B atoms are hydrogenated, respectively. Our findings indicate that the monolayers possess the high $T_{c}$ of 82.0 and 82.5 K, respectively, along with the interesting three-gap superconducting natures. The Fermi sheets showing the obvious three-region distribution characteristics and the abnormally strong electron-phonon coupling (EPC) are responsible for the high-$T_{c}$ three-gap superconductivity. Furthermore, the $T_{c}$ can be dramatically boosted up to 120.0 K under 3.5 \% tensile strain. To a great extent, the high $T_{c}$ is beyond the liquid nitrogen temperature ($77$ K), which is beneficial for the applications in future experiments. This study not only explores the superconducting properties of the monolayers LiBCH and LiCBH, but also offers practical insights into the search for high-$T_{c}$ superconductors.
\end{abstract}
\maketitle

\begin{figure*}[hptb!]
	\centering
	\includegraphics[width=1.0\linewidth]{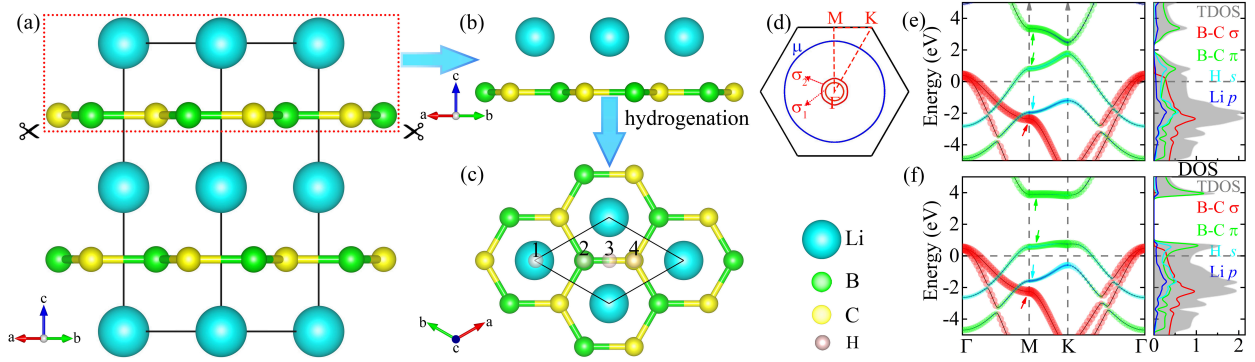}
	\caption{The optimized geometries of (a) bulk LiBC, (b) the exfoliated monolayer LiBC, and (c) the top view of hydrogenated monolayers LiBC with different hydrogenating sites, where the `1, 2, 3, 4' represent the H adatoms located at the top of Li, B, center of B-C bonds, and C atoms, respectively. The unit cell comprising four atoms is marked as a rhombus. (d) Fermi surface of LiBCH and LiCBH. The red and blue curves represent the Fermi sheets originated from the B-C $\sigma$ states (the inner and outer named as $\sigma_{1}$ and $\sigma_{2}$, respectively) and hybridized states of B $\pi$ and H $s$ orbitals (named as $\mu$), respectively. The high symmetric route along $\Gamma$-$M$-$K$-$\Gamma$ is shown in red dotted lines. (e) and (f) represent the electronic properties of monolayers LiBCH and LiCBH, respectively.}
	\label{fig1}
\end{figure*}
$Introduction.$ The multi-gap superconductors generally imply the presence of several superconducting gaps that open up on separate and distinct regions of the Fermi surfaces (FSs) \cite{Suhl}. These materials exhibit a competitive and coupled behavior due to multiple condensates, leading to the emergence of various new physical properties \cite{Milorad,Komendov,Babaev,Garaud,R} and theoretical development \cite{Orlova}. However, the limited availability of multi-gap superconductors poses a significant challenge in confirming these predictions experimentally. This changed after the discovery of two-gap bulk MgB$_{2}$ superconductor \cite{Jun}. According to Bardeen-Cooper-Schrieffer (BCS) theory \cite{Bardeen}, MgB$_{2}$ exhibits a high superconducting transition temperature ($T_{c}$) of $\sim$ 40 K \cite{Jun} due to its high phonon frequency, strong EPC, and large density of states (DOS) at the Fermi level ($E_{F}$). This discovery reignited interest in layered BCS-type compounds that are metallic or can be metalized. Recent discoveries \cite{Pei2022,Lim2022,KAYHAN20121656}, especially superconductivity up to 32 K in the pressurized MoB$_{2}$ \cite{10.1093/nsr/nwad034}, have further fueled this renaissance. It is also worth noting that a strong covalent $\sigma$ bond above the $E_{F}$, consisting of a spin singlet pair with opposite spins \cite{Gao}, is essential for superconductivity. It breaks the binding force of the band, allowing electrons to become itinerant and resulting in strong EPC and a potentially high $T_{c}$ \cite{Gao,Hyoung,An,Kong}. In the pursuit of identifying novel phonon-mediated superconductors with higher $T_{c}$ under ambient pressure, several potential candidates have been proposed.

Bulk LiBC \cite{Michael} has garnered attention for its analogical resemblance and straightforward fabrication. It is isostructural to MgB$_{2}$ \cite{Ravindran}, which effortlessly meets the aforementioned high-$T_{c}$ criteria. The crystal structure is derived from the fully intercalated graphite structure of MgB$_{2}$: Mg $\rightarrow$ Li and B$_{2}$ $\rightarrow$ BC, resulting in B being the nearest neighbor to C due to the hexagonal arrangement of BC layers. Although Li atom has one less valence electron than Mg, C possesses one more electron than B. This renders bulk LiBC isovalent with MgB$_{2}$ and hugely probable to exhibit metallic properties. Unfortunately, it is a semiconductor with covalent bonding states situated on the top of the valence bands \cite{Michael,P}. 

Over the years, extensive efforts have been dedicated to transforming LiBC into a metallic state in order to explore its superconducting properties. Currently, there are several approaches to metallize it: (1) It was theoretically discovered that the injection of Li vacancies (Li$_{0.5}$BC) in bulk LiBC can raise the $\sigma$-bonding states above $E_{F}$ and result in a superconducting state with a $T_{c}$ of approximately 100 K \cite{Rosner}, while a higher concentration of Li vacancies in Li$_{0.125}$BC was predicted to yield a $T_{c}$ of up to 65 K \cite{Dewhurst}. However, with the introduction of Li vacancies, strong lattice distortion follows \cite{Fogg}, causing a significant change in the band structure and weakening the positive effects of lifting the $\sigma$-bonding band above the $E_{F}$ \cite{Fogg}. Consequently, superconductivity have never been observed in synthetic Li$_{x}$BC \cite{Fogg,A,Dmitri,Fogg2}; (2) To suppress the impact of introducing the lattice distortion and simultaneously metalize LiBC-type compounds, Miao \cite{Miao} and Gao \cite{Gao} $et$ $al.$ proposed an idea of partial substitution between C and B atoms. Investigations have demonstrated that certain LiBC-type compounds, such as LiB$_{1.1}$C$_{0.9}$ \cite{Miao}, Li$_{3}$B$_{4}$C$_{2}$ \cite{Gao}, LiBC$_{3}$ \cite{PhysRevB.102.144504}, and Li$_{4}$B$_{5}$C$_{3}$ \cite{Bazhirov}, exhibit superconductivity with $T_{c}$ values of 36, 50, and 16.8 K, respectively. However, the synthesis of these B-enriched stoichiometric compounds has proven to be challenging \cite{V}. Additionally, the hole-doped Li$_{x}$B$_{1.15}$C$_{0.85}$ exhibits a significant decrease in resistivity below 20 K but does not exhibit superconductivity \cite{Ayako}; (3) Another possible way to realize the metalization is applying pressure. However, Lazicki \textit{et al}. found that even at 60 GPa \cite{Lazicki}, the crystal structure of LiBC remains unaltered. Subsequently, the theoretically calculated pressure required for its metalization is up to 345 GPa. However, pressure also eliminates the similarity of electronic structure between LiBC and MgB$_{2}$ \cite{Lazicki,Meiguang}, resulting a new phase; (4) Dimensional reduction is a significant route to achieve metalization of bulk LiBC. Gao and Modak $et$ $al.$ have investigated the effects of dimensional reduction on metalization, leading to superconductivity with $T_{c}$ of $\sim$90 K in trilayers LiB$_{2}$C$_{2}$ \cite{Gao2020} and 70 K monolayer LiBC \cite{Modak}, respectively. This method seems to be a viable way to explore the superconductivity of LiBC compound, but experimental verification is imperative.

Is there another possible approach to realize the metalization of LiBC? Here, one initial idea that comes to our mind is to explore hydrogenation as a means to achieve it. Dense metallic hydrogen \cite{Ashcroft} can lead to a high-temperature superconductor with $T_{c}$ of 242 K \cite{Cudazzo} under ultra-high pressure ($\sim$ 400 GPa) \cite{PhysRevLett.112.165501,PhysRevLett.114.105305}, accompanying with multi-gap superconducting characteristics \cite{Cudazzo}. Over the years, hydrogenation, as an active way to boost the $T_{c}$ of superconductors, have been widely studied \cite{PhysRevLett.123.077001,Yang_2023,PhysRevB.105.245420,PhysRevMaterials.6.054003,Li2022,HAN2023100954,doi:10.1021/jacs.2c05683,D2NR01939F}. Thus, the idea to achieve the metalization of LiBC compound is hydrogenation. First, monolayer LiBC can be exfoliated from the bulk LiBC \cite{Michael} to realize dimensional reduction. Subsequently, H atom can be added to the side of Li layer for realizing its hydrogenation. In this work, presenting the common advantages of dimensional reduction \cite{Gao2020,Modak,PhysRevB.96.094510,Liu2023}. and hydrogenation \cite{Ashcroft,Cudazzo}, LiBCH and LiCBH are predicted to be high-$T_{c}$ superconductors.

All calculations reported in this work were performed using density functional theory. We employed the QUANTUM ESPRESSO (QE) package \cite{Paolo} for the electronic structure, lattice dynamics, and the EPW code \cite{Giustino,Jesse,Margine} for the EPC and superconducting gaps. More computational details are presented in the Supplementary Materials (SM) \cite{SM}.


$Results$ $and$ $discussion.$ The starting point is to ascertain the equilibrium structure of the hydrogenated LiBC system. To achieve this goal and eliminate the influence of different H atomic concentrations, we examine several potential high-symmetry sites shown in Fig. 1(c), and simulate the corresponding dynamic stabilities. As detailed in Fig. S1 of SM \cite{SM}, only two dynamically stable structures with one H adatom per unit cell are observed. Their sufficiently optimized structures are distinguished by H sites in which B and C are hydrogenated, named as LiBCH and LiCBH, respectively. The total energy per atom of LiBCH is lower about 20 meV than that of LiCBH. The calculated crystal parameters can be obtained in Table S2 of SM \cite{SM}. Moreover, stability is important for a new material. Thus, we prove their thermodynamic, mechanical, dynamic, and electronic stability, where the details are provided in the Section E of SM \cite{SM}.

Based on the overall analysis in Section F of SM \cite{SM}, we can draw the following conclusion about charge transfer: Li-$s$ and C/B-$p_{z}$ orbitals lose charges; Li-$p$, C/B-$p_{x}$, $p_{y}$, and H-$s$ orbitals get charges. On the whole, C/H and B/Li atoms obtain and lose charges, respectively. The band structures and DOS for LiBCH and LiCBH are shown in Figs. 1(e) and 1(f). The orbitals with little contributions at the $E_{F}$, such as the $s$ orbitals of B(C), are omitted. There are three bands cross the $E_{F}$, one band along the high-symmetry points $M$ and $K$ is mainly composed of the hybridized states of B/C-$\pi$ and H-$s$ orbitals, two bands around the BZ center originate from the B/C-$\sigma$ states, leading to two hole pockets. It is indicative of metalization motivated by covalent states, analogous to what is observed in other multi-gap superconductors MgB$_{4}$ \cite{PhysRevB.101.104507}, AlB$_{2}$-based films \cite{PhysRevB.100.094516}, MgB$_{2}$ \cite{An,Bazhirov,PhysRevLett.86.4656,PhysRevLett.87.037001}. Moreover, the band gaps between the valence band maximum (VBM) and conduction band minimum (CBM) of LiBCH and LiCBH around $K$ point are 0.722 and 3.073 eV, respectively. As depicted in the DOS, the dominant electronic states near the $E_{F}$ come from the contribution of B/C-$\sigma$, -$\pi$ and H-$s$ orbitals, with minor contribution from Li-$p$ orbitals. Hence, compared with those electrons of LiBC only from B/C-$\sigma$ and -$\pi$ states \cite{Modak}, hydrogenation can bring higher electronic DOS at the $E_{F}$, which benefits EPC. However, Li-$p$ orbitals play essential roles in the construction of more localized Wannier functions, as shown in Fig. S2 and Table S1 of SM \cite{SM}. The constraint of B-C lattice subjects the Li atom to substantial lattice chemical pressure with energetic level transitions from the $1s$ to $2p$ orbitals. This feature is not unique and has also been reported in monolayer LiBC \cite{Modak}. The van Hove singularities are indicated by colored arrows, with each color representing a different dominant orbital. Moreover, the distribution behaviors, prompting us to employ the anisotropic Migdal-Eliashberg theory, of atomic orbitals on the FSs and detailed phonon vibrations are provided in the SM \cite{SM}.

\begin{figure}[hptb!]
	\centering
	\includegraphics[width=1\linewidth]{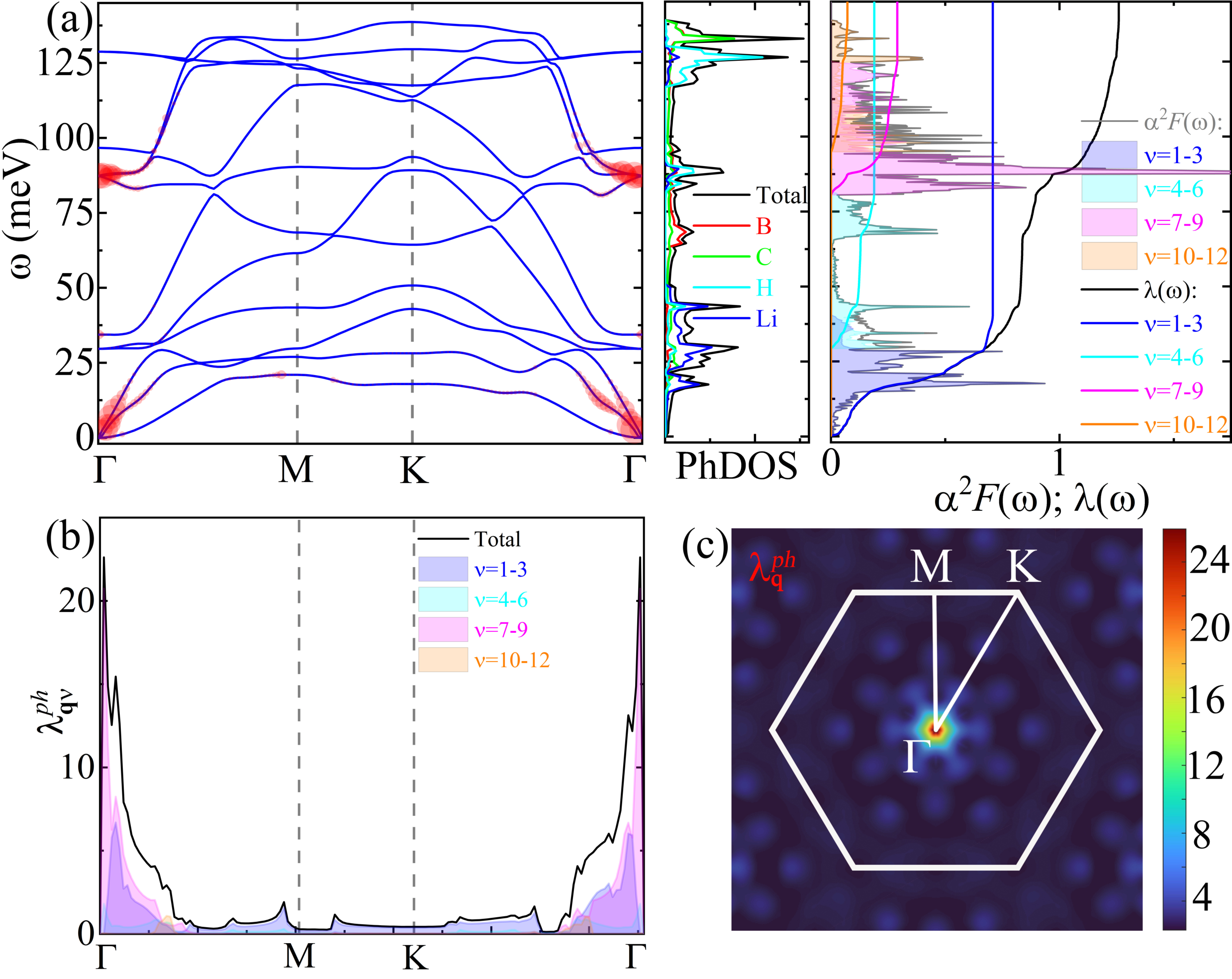}
	\caption{Properties of lattice dynamics and EPC parameters $\lambda^{\textit{ph}}_{\textbf{q}}$ for the LiBCH monolayer. (a) Phonon weighted with the EPC parameters $\lambda^{\textit{ph}}_{\textbf{q}}$ (red), PhDOS, and the total cumulative EPC $\lambda(\omega)$=2$\int\alpha^{2}F(\omega)$/$\omega$$d\omega$ with the corresponding mode-resolved Eliashberg spectral function $\alpha^{2}F(\omega)$. (b) EPC $\lambda^{\textit{ph}}_{\textbf{q}\nu}$ along the high-symmetry line $\Gamma$-$M$-$K$-$\Gamma$. (c) EPC $\lambda^{\textit{ph}}_{\textbf{q}}$ projected in the whole BZ.}
	\label{fig2}
\end{figure}
Subsequently, we concentrate on the superconductivity of LiBCH and LiCBH. The related superconducting properties are shown in Figs. 2 and S10 for LiBCH and LiCBH, respectively. As shown in Fig. 2(a), the strong projections of EPC parameters $\lambda^{\textit{ph}}_{\textbf{q}}$ are located in two regions: one is the acoustic phonon modes around $\Gamma$ point; the other is the BZ-center optical $E$ modes originating from the in-plane (seen in Table S6 and Figs. S9 and S11 of SM \cite{SM}) B-C stretching. The strong coupling between the B/C-$\sigma$ states and the in-plane phonon vibration of B-C graphene-like net results in the large $\lambda^{\textit{ph}}_{\textbf{q}\nu}$ (Fig. 2(b)) around $\Gamma$ point and leads to some high peaks of Eliashberg functions $\alpha^{2}F(\omega)$, particularly the maximum at the optical $E$ modes points. The anomalously large EPC strengths of $\lambda^{\textit{ph}}_{\textbf{q}\nu}$ are from the dramatic softening of the optical $E$ modes with the existence of Kohn anomaly \cite{PhysRevLett.2.393}, rather than the effect of nesting function shown in Fig. S5 of SM \cite{SM}. There is a strong concordance between the regions where optical phonons experience softening in the BZ and the diameter of hole Fermi sheets that arise from the $\sigma$-bonding B/C orbitals, demonstrating a typical attribute associated with the Kohn effect \cite{PhysRevB.101.104507,PhysRevB.104.174519}. These findings are further corroborated by the mode-resolved $\alpha^{2}F(\omega)$ and $\lambda(\omega)$. Moreover, for LiBCH (LiCBH), the three low-frequency acoustic phonon branches $\nu$=1-3 contribute 56.3 (62.5) \% of the total EPC $\lambda(\omega)$, while the remaining modes $\nu$=4-6, 7-9, and 10-12 contribute 15.0 (15.0) \%, 23.0 (18.6) \%, and 5.7 (3.9) \%, respectively. The total EPC $\lambda$ is up to 1.26 (1.36).

\begin{figure}[hptb!]
	\centering
	\includegraphics[width=1\linewidth]{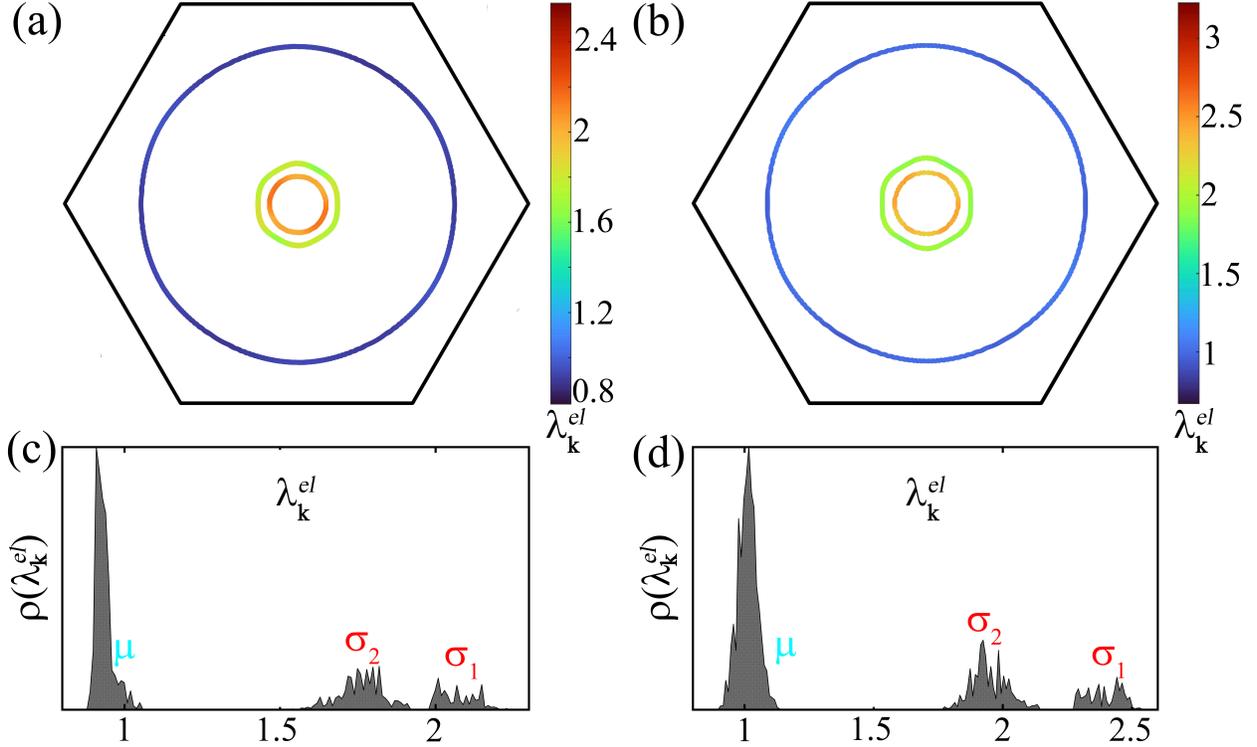}
	\caption{The EPC parameters $\lambda^{\textit{el}}_{\textbf{k}}$ and strength distribution $\rho(\lambda^{\textit{el}}_{\textbf{k}})$ for (a), (c) LiBCH and (b), (d) LiCBH, respectively.}
	\label{fig3}
\end{figure}

The EPC parameters $\lambda^{\textit{el}}_{\textbf{k}}$ on the FS and strength distribution $\rho({\lambda^{\textit{ph}}_{\textbf{k}}})$ are shown in Fig. 3. The detailed discussion about the differences and correlations can be obtained in Ref. \cite{RevModPhys.89.015003}. For LiBCH (LiCBH), the range of whole EPC $\lambda^{\textit{el}}_{\textbf{k}}$ is from 0.87 (0.85) to 2.31 (2.63). The latter is slightly larger than that of the former. Moreover, the larger EPC $\lambda^{\textit{el}}_{\textbf{k}}$ exhibits in the $\sigma_{1}$ and $\sigma_{2}$ sheets, while $\mu$ sheet possesses the smaller value of $\lambda^{\textit{el}}_{\textbf{k}}$. The strength distributions of $\sigma_{1}$, $\sigma_{2}$, and $\mu$, namely $\rho(\lambda^{\textit{el},\sigma_{1}}_{\textbf{k}})$, $\rho(\lambda^{\textit{el},\sigma_{2}}_{\textbf{k}})$, and $\rho(\lambda^{\textit{el},\mu}_{\textbf{k}})$, are from 1.98 (2.27), 1.47 (1.67), and 0.87 (0.85) to 2.31 (2.63), 1.92 (2.14), and 1.07 (1.18), respectively. The resulting average of EPC $\lambda^{\textit{el}}$ is 1.69 (1.82). According to the aforementioned analysis, we can find that the high-frequency optical phonon of the dramatic softening around $\Gamma$ point, e.g., the $E$ modes, and the acoustic phonon significantly contribute to forming the large electronic EPC $\lambda^{\textit{el}}_{\textbf{k}}$ on the $\sigma_{1}$ and $\sigma_{2}$ sheets. 

\begin{figure*}[hptb!]
	\centering
	\includegraphics[width=1\linewidth]{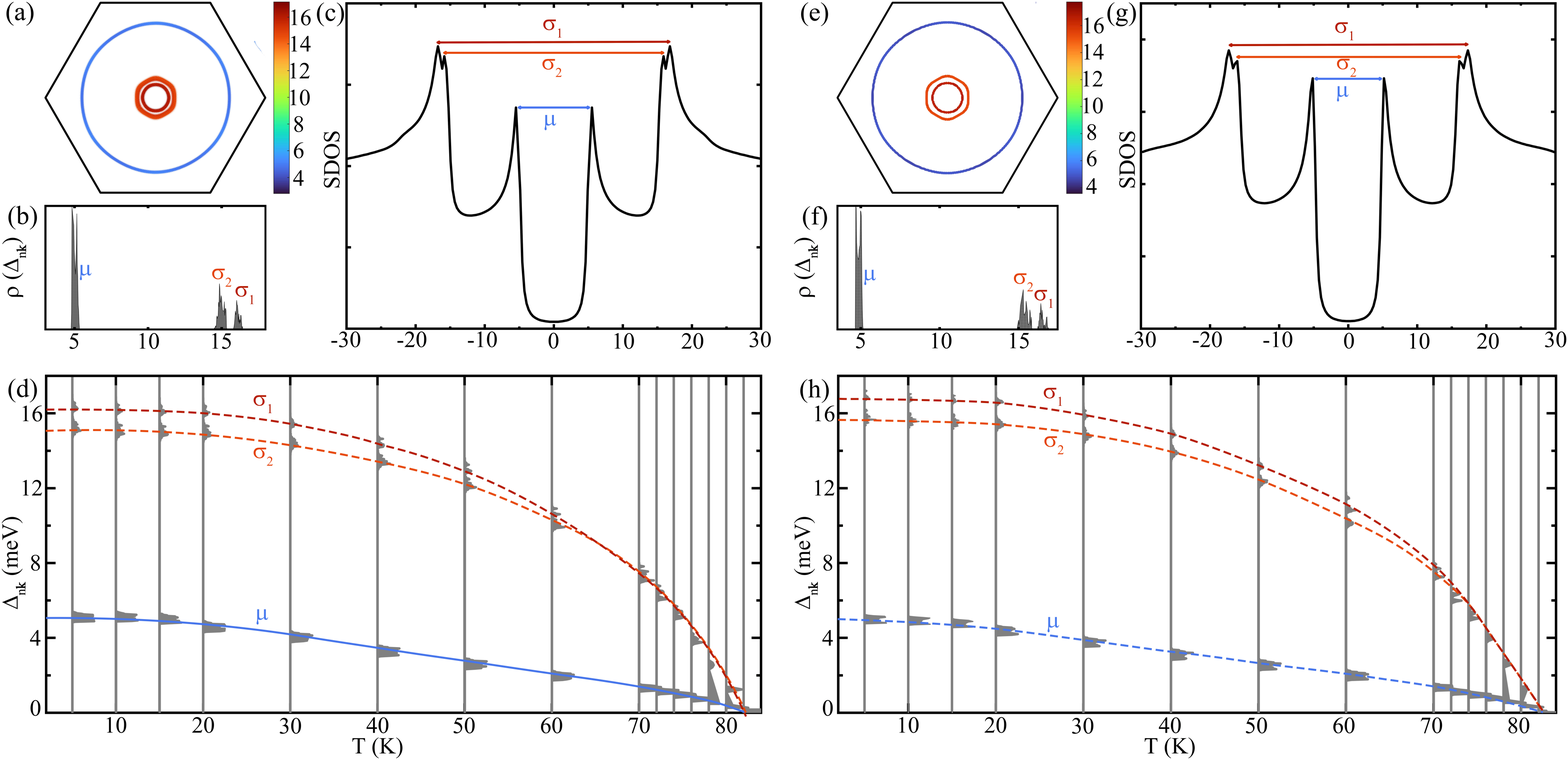}
	\caption{Superconducting gap properties of LiBCH and LiCBH. (a) Momentum-resolved superconducting gap $\Delta_{n\textbf{k}}(\omega=0)$ on the FS at $T$ = 5 K; (b) the strength distributions of gaps $\rho(\Delta_{n\textbf{k}})$; (c) SDOS at $T$ = 5 K; (d) energy distribution of the gaps $\Delta_{n\textbf{k}}$ versus $T$ for LiBCH. (e) $\sim$ (h) The corresponding superconducting gaps results for LiCBH. The dark red, red, and blue lines in (d) and (h) are the eye guides for the gaps $\Delta_{\sigma_{1}}$, $\Delta_{\sigma_{2}}$, and $\Delta_{\mu}$, respectively. (Note: the heights of the histograms are multiplied by a factor of 2.5 while plotting for visibility in (d) and (h).)}
	\label{fig4}
\end{figure*}

Meanwhile, the different regions with different electronic states on the FS can lead to the separation of strength distributions ($\lambda^{\textit{el}}_{\textbf{k}}$) between the $\sigma$ and $\mu$ sheets. As demonstrated in Figs. 3(c) and 3(d), the behavior that $\sigma$ sheets are divided into $\sigma_{1}$ and $\sigma_{2}$ sheets results in the formation of three-region strength distribution characteristics of $\lambda^{\textit{el}}_{\textbf{k}}$. The reason for separation between the strength distributions can be attributed to the extra slight distribution of the phonon EPC $\lambda^{\textit{ph}}_{\textbf{q}}$, e.g., between $\Gamma$ and $K$ points with a proximity bias towards $\Gamma$ point, shown in Figs. 2(c) and S10(c). These slight distributions of $\lambda^{\textit{ph}}_{\textbf{q}}$ originate from the weak phononic softening located around the $\Gamma$ point, shown in Figs. 2(a) and S10(a). The same results have also been observed in MgB$_{4}$ \cite{PhysRevB.101.104507} and AlB$_{4}$ \cite{PhysRevB.100.094516}.


Based on fully anisotropic Migdal-Eliashberg theory \cite{migdal1958interaction,eliashberg1960interactions,Philip,Margine,Marsiglio}, we now turn to analyze the superconducting gaps, shown in Fig. 4. Combing with Fig. 3, it is worth noting that the regions with larger gaps on FSs shown in Figs. 4(a) and 4(e) coincide with those with stronger EPC $\lambda^{el}_{\textbf{k}}$, with the strongest gap $\Delta_{\sigma_{1}}$, secondary gap $\Delta_{\sigma_{2}}$, and hybridized-states-included weakest gap $\Delta_{\mu}$ opening on the $\sigma_{1}$, $\sigma_{2}$, and $\mu$ sheets, respectively. The superconducting DOS (SDOS) can be calculated with the formula S24. As shown in Figs. 4(c) and 4(g), the SDOS with three distinguished peaks corresponding to the $\Delta_{\sigma_{1}}$, $\Delta_{\sigma_{2}}$, and $\Delta_{\mu}$ further demonstrates the nature of three superconducting gaps. For LiBCH (LiCBH) at the temperature $T$ = 5 K, shown in Figs. 4(b) and 4(f), the ranges of three gaps $\Delta_{\sigma_{1}}$, $\Delta_{\sigma_{2}}$, and $\Delta_{\mu}$ are 15.75 (16.20) $\sim$ 16.65 (17.10), 14.46 (14.91) $\sim$ 15.36 (15.87), and 4.80 (4.68) $\sim$ 5.40 (5.19) meV, respectively. The largest and smallest gap divisions originating from the $\Delta_{\sigma_{2}}$, $\Delta_{\mu}$ and $\Delta_{\sigma_{2}}$, $\Delta_{\sigma_{1}}$ are 9.06 (9.72) and 0.39 (0.33) meV, respectively. These results agree well with three-region distribution characteristics of electronic EPC $\lambda^{\textit{el}}_{\textbf{k}}$, indicating that LiBCH and LiCBH are evidently three-gap superconductors. As depicted in Figs. 4(d) and 4(h), the typical BCS-type temperature dependence occurs due to the changes in gaps $\Delta_{n\textbf{k}}$ versus $T$. Moreover, the value of $\Delta_{\mu}$ is enlarged from $\sim$2.7 meV of LiBC \cite{Modak} to $\sim$5 meV by introducing the hydrogen. As the temperature increases, the values of $\Delta_{n\textbf{k}}$ gradually vanish and converge at 82.0 and 82.5 K for LiBCH and LiCBH, respectively. Their $T_{c}$s are higher than 17 $\sim$ 47 K of AlB$_{2+x}$ films \cite{PhysRevB.100.094516}, 52 K of MgB$_{4}$ \cite{PhysRevB.101.104507}, 53 K of MgB$_{2}$ with the strain of 4.5\% \cite{PhysRevB.96.094510}, 67 K of hydrogenated MgB$_{2}$ \cite{PhysRevLett.123.077001}, and the pristine $T_{c}$s of carbon-cage network \cite{Hai2023}. It is interesting that both $T_{c}$s are beyond the liquid nitrogen temperature (77 K), which are highly significant and evidently beneficial for the application in the future experiments. 

Here, we draw a short discussion about superconducting parameters calculated by the Allen-Dynes (AD) formula. For LiBCH (LiCBH), the results suggest that the EPC $\lambda_{AD}$ and $T_{c}$ are 1.19 (1.33) and 36.71 (32.71) K, respectively, about 55.2 (60.4) \% less than the values analyzed from the multi-band theory. This clearly substantiates that the presence of multigaps, three gaps, here plays a crucial role in capturing the rising $T_{c}$.


Due to the significant enhancement of superconductivity under the biaxial strain in numerous 2D materials, e.g., hydrogenated MgB$_{2}$ \cite{PhysRevLett.123.077001}, monolayer MgB$_{2}$ \cite{PhysRevB.96.094510}, hole-doped graphene \cite{PhysRevLett.196802} and graphane \cite{PhysRevLett.105.037002}, its effect on the superconductivity of LiBCH and LiCBH should be also explored. Under strains, the out-of-plane vibration of H atoms gradually plays an essential role in the formation of larger EPC and stronger softening modes, meaning that H atoms can significantly enhance the EPC. As shown in Figs. S14(a) and S14(d) of SM \cite{SM}, the regions with lager $\lambda^{\textit{el}}_{\textbf{k}}$ on the FS appear on the $\mu$ sheets, rather than the $\sigma$ sheets of equilibrium cases ($\varepsilon$ = 0\%). Meanwhile, the EPC constants $\lambda$ for LiBCH and LiCBH are sharply boosted up to 2.54 and 3.45, respectively, leading to stronger EPC. Compared with LiBC \cite{Modak}, it demonstrates the importance of H atoms to the boost of superconductivity. Subsequently, their $T_{c}$s are simultaneously boosted up to 103.5 and 120.0 K, respectively. It is incredibly interesting that these predicted high $T_{c}$s beyond 100 K at ambient pressure are rare in the current researches of 2D superconductors. More detailed information about charge density wave (CDW) transition and other superconducting parameters, e.g., $\lambda^{\textit{ph}}_{\textbf{q}}$,  can be shown in Fig. S12 - S14 of SM \cite{SM}.

Before drawing a conclusion, it is extremely worth noting that this research represents a direction for searching high-$T_{c}$ superconductors at ambient pressure. Taking full advantage of the following aspects: the nature of multigaps, the $\sigma$ bonds of B-C graphene-like net, and H atoms. The multigaps can possibly open larger superconducting gaps, resulting in the higher $T_{c}$ \cite{PhysRevLett.94.037004,PhysRevB.75.054508,PhysRevB.101.104507}. The $\sigma$ bonds of B-C graphene-like net usually provides stronger EPC parameters $\lambda^{\textit{ph}}_{\textbf{q}}$ which can trigger higher peaks of $\alpha^{2}F(\omega)$ and subsequently stronger EPC \cite{Gai}. The advantages of H atoms can rise the upper limit of phonon frequency $\omega$ and boost the value of $T_{c}$, especially in the strain engineering. In addition, dimensional reduction can also effectively increase $T_{c}$ \cite{PhysRevB.96.094510,Liu2023}. Based on these principles, we believe that more superconductors with higher $T_{c}$ can be discovered. 

In summary, we find two stable hydrogenated LiBC configurations from LiBCH$_{x}$($x$=1, 2, 3, 4) with different H atomic concentrations. Subsequently, their stabilities, charge transfer, and electronic/phononic properties are studied. By solving the fully anisotropic Migdal-Eliashberg equations, superconducting behaviors are investigated systematically. We observe an interesting nature of three-gap superconductivity with high $T_{c}$s of 82.0 and 82.5 K, and reveal that the strong electronic EPC parameters $\lambda^{\textit{el}}_{\textbf{k}}$ with evident three-region distribution characteristics are responsible for this high-$T_{c}$ three-gap superconductivity. Moreover, our calculations suggest that the dominating phononic EPC parameters $\lambda^{\textit{ph}}_{\textbf{q}}$, resulted from the acoustic and high-energy optical softening phonon modes around the BZ center, contribute to larger $\lambda^{\textit{el}}_{\textbf{k}}$ between the $\sigma_{1}$ and $\sigma_{2}$ Fermi sheets. The distinct separation of the $\sigma$ sheets further indicate the formation of three-region distribution characteristics of $\lambda^{\textit{el}}_{\textbf{k}}$ and the robust three-gap superconducting behaviors. We also discover that the coupling between the covalent metalized states and in-plane phonon vibration modes originating from B-C graphene-like net play an essential role in superconductivity. In addition, the $T_{c}$s can be dramatically boosted up to 103.5 and 120.0 K under the small strains $\varepsilon$ = 2.1 \% and 3.5 \%, respectively. The large boost of $T_{c}$ originates from the H atomic out-of-plane phononic vibrations on the outer $\mu$ sheet, rather than the inter $\sigma$ sheets in the pristine cases. This work not only explores the superconducting properties of the LiBCH and LiCBH monolayers but also provides practical insights into the search for high-$T_{c}$ superconductors.

$Acknowledgements.$ This work was supported by the National Natural Science Foundation of China (Grant Nos. 12175023, 12074213, 11574108), the Major Basic Program of Natural Science Foundation of Shandong Province (Grant No. ZR2021ZD01), and the Project of Introduction and Cultivation for Young Innovative Talents in Colleges and Universities of Shandong Province.

\end{document}